\documentclass[conference]{IEEEtran}
\ifCLASSINFOpdf
\else
\fi
\usepackage{amsmath,amssymb}
\usepackage[dvipdfmx]{color}
\usepackage[dvipdfmx]{graphicx}
\usepackage{amsthm}
\theoremstyle{definition}

\newtheorem*{theorem*}{Theorem}

\usepackage{txfonts}
\usepackage{setspace}
\begin{document}
%
\title{SNR Maximization as a Non-Linear Programming \\-Towards Optimal Spreading Sequence-}

\author{\IEEEauthorblockN{Hirofumi Tsuda}
\IEEEauthorblockA{Department of Applied Mathematics and Physics\\
Graduate School of Informatic\\ Kyoto University\\
Kyoto, Japan\\
Email: tsuda.hirofumi.38u@st.kyoto-u.ac.jp}
\and
\IEEEauthorblockN{Ken Umeno}
\IEEEauthorblockA{Department of Applied Mathematics and Physics\\
Graduate School of Informatic\\ Kyoto University\\
Kyoto, Japan\\
Email: umeno.ken.8z@kyoto-u.ac.jp}}


%


\maketitle

\begin{abstract}
Signal to Noise Ratio (SNR) is an important index for wireless communications. There are many methods for increasing SNR. In CDMA systems, spreading sequences are used. We consider the frequency-selective wide-sense-stationary uncorrelated-scattering (WSSUS) channel and evaluate the worst case of SNR. We construct the non-linear programing for maximizing the lower bound of the average of SNR. This problem becomes the convex programming if we did not take into account the norm constraint. We derive necessary conditions for optimal spreading sequences for the problem. 
\end{abstract}


%
\IEEEpeerreviewmaketitle

\section{Introduction}
Communications in mobile radio networks are necessary for our lives. Multiple access \cite{multiple} realizes that many people can communicate each other at the same time. For realizing a multiple access, there are some techniques to communicate, for example, frequency-division multiple access (FDMA), time-division multiple access (TDMA) and code division multiple access (CDMA) \cite{dscdma}. \\
In particular, CDMA is used for the 3G mobile communication system. In CDMA systems, spreading sequences are utilized as codes. On the other hand, OFDM has been recently used for 4G and broadband WiFi systems. Since the number of users has been dramatically increased and each user has not only one device, it is necessary to multiplex to communicate among a number of devices in 5G. In using the current technology such as OFDM, it is difficult to increase the capacity \cite{shannon} since the frequency spectrum band which we can use is limited. To achieve high capacity, we focus on CDMA again.\\
In general, high Signal to Noise Ratio (SNR) has been demanded for achieving high spectral efficiency \cite{efficiency}. In an asynchronous access environment, to increase SNR, many methods have been proposed, for example \cite{mmse} and \cite{ml}.\\
 In CDMA, crosscorrelation is treated as a basic component of interference noise, and autocorrelation is related to code synchronization at receiver side and the fading noise. It is necessary to reduce crosscorrelation to achieve high capacity and it is desirable that the second peak in autocorrelation is low for code synchronization and fading noise.\\
The current spreading sequence of 3G CDMA systems is the Gold code \cite{gold}. It is known that this spreading sequence as well as the Kasami sequence is optimal in all the binary spreading sequences. In \cite{chaos_cdma} and \cite{chaos_mod}, the use of chaotic spreading sequences has been proposed. These spreading sequences are obtained from chaotic maps. Examples of such spreading sequences are \cite{logistic} \cite{ergotic} \cite{autocor} \cite{mazzini}. For these chaos-based DS-CDMA systems, the performance in fading channels has been investigated in \cite{kaddoum} and \cite{multipath}. By taking this approach, the optimal sequences are not obtained in SNR.\\
On the contrary, our approach is to improve spreading sequences for increasing SNR in a general setting where explicit form of generating spreading sequence is not assumed. We derive spreading sequences to maximize SNR. For CDMA systems, the expressions of SNR have been obtained in \cite{mazzini} and \cite{pursley}. However, the denominators of these expressions are not convex functions and differentiations of them are complicated. Therefore, it is difficult to maximize SNR when we consider the spreading sequences as parameters. In our analysis in the following sections, we make some assumption about the Gaussian process and show the new expressions of SNR whose denominator is a convex function. From this expression, we construct the non-linear programming for maximizing the lower bound of SNR and obtain the necessary conditions for optimal spreading sequences.

\section{Asynchronous CDMA Model}
In this section, we fix our model and mathematical symbols that will be used in the following sections. We consider the following asynchronous CDMA model \cite{pursley}\cite{borth}. Let $N$ be the length of spreading sequences. The user $k$'s data signal $b_k(t)$ is expressed as
\begin{equation}
b_k(t) = \sum_{n=-\infty}^{\infty} b_{k,n} p_{T}(t - nT),
\end{equation}
where $b_{k,n} \in \{-1,1\}$ is the $n$-th component of bits which user $k$ send, $T$ is the duration of one symbol and $p_{T}(t)$ is a rectangular pulse written as
\[p_T(t) = \left\{ \begin{array}{c c}
1 & 0 \leq t \leq T\\
0 & \mbox{otherwise}
\end{array} \right. .\]
The user $k$'s code waveform $s_k(t)$ is expressed as
\begin{equation}
s_k(t) = \sum_{n=-\infty}^{\infty} s_{k,n} p_{T_c}(t - nT_c),
\end{equation}
where $s_{k,n}$ is the $n$-th component of the user $k$'s spreading sequence and $T_c$ is the width of the each chip which satisfies $NT_c = T$. The sequence $s_{k,n}$ has the period $N$, that is, $s_{k,n}=s_{k,n+N}$. 
The user $k$'s transmitted signal $\zeta_k(t)$ is
\begin{equation}
\zeta_k(t) = \sqrt{2P} \operatorname{Re}[s_k(t)b_k(t)\exp(j \omega_c t + j\theta_k)],
\label{eq:carrer}
\end{equation}
where $P$ is the common signal power, $\omega_c$ is the common carrier frequency and $\theta_k$ is the phase of the $k$-th user.\\
We consider a Rician fading channel. In this channel model, the received signal $r(t)$ is
\begin{equation}
r(t) = \sum_{k=1}^K \operatorname{Re}\left[u_k(t - \tau_k)\exp\left(j \omega_c t + j \psi_k \right) \right] + n(t),
\end{equation}
where $\tau_k$ is the time delay, $\psi_k = \theta_k - \omega_c \tau_k$, $n(t)$ is the additive white Gaussian noise (AWGN), and $u_k(t)$ is
\begin{equation}
u_k(t) = \gamma_k \int_{-\infty}^{\infty} h_k(\tau, t)x_k(t-\tau)d\tau + x_k(t),
\label{eq:rice}
\end{equation}
\begin{equation}
x_k(t) = \sqrt{2P} s_k(t)b_k(t).
\label{eq:rice2}
\end{equation}
The first term of Eq. (\ref{eq:rice}) is the component of a faded signal and the second one is the component of a direct wave. The function $h_k(\tau, t)$ is the zero-mean complex Gaussian random process and $\gamma_k$ is the nonnegative real parameter which represents the transmission coefficient for the user $k$'s signal. If the received signal $r(t)$ is the input to a correlation receiver matched to $\zeta_i(t)$, then the corresponding output $Z_i$ is
\begin{equation}
Z_i = \int_{0}^T r(t) \operatorname{Re}[s_i(t-\tau_i)\exp(j \omega_c t + j \psi_i)]dt.
\label{eq:output}
\end{equation}
Without loss of generality, we assume $\tau_i = 0$ and $\theta_i = 0$ and hence $\psi_i = 0$. With a low-pass filter, we can ignore double frequency terms, and rewrite Eq. (\ref{eq:output}) as
\begin{equation}
\begin{split}
Z_i &= \frac{1}{2} \sum_{k=1}^K \int_{0}^T \operatorname{Re}[u_k(t- \tau_k)\overline{s_i(t)}\exp(j \psi_k)]dt\\
&+ \int_0^T n(t)\operatorname{Re}[s_i(t)\exp(j \omega_c t)],
\label{eq:output2}
\end{split}
\end{equation}
where $\overline{z}$ is complex conjugate of $z$ and 
\begin{equation}
\overline{s_i(t)} = \sum_{n=-\infty}^{\infty} \overline{s_{k,n}} p_{T_c}(t - nT_c).
\end{equation}
In obtaining  Eq. (\ref{eq:output2}), we have used the identity
\begin{equation}
2\operatorname{Re}[z_1]\operatorname{Re}[z_2] = \operatorname{Re}[z_1z_2] + \operatorname{Re}[z_1\overline{z_2}],
\label{eq:thm1}
\end{equation}
where $z_1, z_2 \in \mathbb{C}$.\\
Similar to \cite{pursley}, we assume that the phase $\psi_k$, time delays $\tau_k$ and bits $b_{k,n}$ are independent random variables and they are uniformly distributed on $[0, 2\pi)$, $[0,T)$ and $\{-1 ,1\}$. Without loss of generality, we assume that $b_{i,0} = +1$.\\
To evaluate SNR, we define
\begin{equation}
\mu_{i,k}(\tau; t) = b_k(t - \tau)s_k(t - \tau)\overline{s_i(t)}
\end{equation}
and
\begin{equation}
\xi_{i, k_1,k_2}(\tau_1, \tau_2; t_1, t_2) = \mu_{i, k_1}(\tau_1; t_1) \overline{\mu_{i, k_2}}(\tau_2; t_2).
\end{equation}
For convenience, we write $\mu_{i,i}$ and $\xi_{i,i,i}$ as $\mu_i$ and $\xi_i$. We divide $Z_i$ into the four signals, the user $i$'s desired signal $D_i$, the user $i$'s faded signal $F_i$, the interference signal $I_i$ and the AWGN signal $N_i$. They are expressed as
\begin{equation}
\begin{split}
D_i &= \sqrt{\frac{P}{2}} \int_0^T b_i(t)dt\\
F_i &= \sqrt{\frac{P}{2}}\operatorname{Re}[\tilde{F}_i]\\
I_i &= \sqrt{\frac{P}{2}} \sum_{\substack{k=1 \\ k \neq i}}(\operatorname{Re}[\gamma_k \tilde{I}_{i,k}] + \operatorname{Re}[\tilde{I}'_{i,k}])\\
N_i &= \int_0^Tn(t)\operatorname{Re}[s_i(t)\exp(j\omega_ct)]
\end{split}
\label{eq:signals}
\end{equation}
where
\begin{equation*}
\begin{split}
\tilde{F}_i & = \int_0^T \int_{-\infty}^{\infty}\gamma_i h_i(\tau,t) \mu_i(\tau ; t) d\tau dt\\
\tilde{I}_{i,k} &= \int_0^T \int_{-\infty}^{\infty} h_k(\tau,t - \tau_k) \mu_{i,k}(\tau_k + \tau ; t) \exp(j \psi_k) d \tau d t\\
\tilde{I}'_{i,k} &= \int_0^T \mu_{i,k}(\tau_k; t) \exp(j \psi_k) dt.\\
\end{split}
\end{equation*}
From these expressions, we decompose $Z_i$ as
\begin{equation}
Z_i = D_i + F_i + I_i + N_i.
\end{equation}
\section{Evaluation of SNR}
Since $\operatorname{E}\{F_i\} = \operatorname{E}\{N_i\} = \operatorname{E}\{N_i\} = \operatorname{E}\{N_i\} = 0$ and $\displaystyle\operatorname{E}\{D_i\} = T\sqrt{P/2}$, then $\displaystyle \operatorname{E}\{Z_i\} = T\sqrt{P/2}$, where $\operatorname{E}\{X\}$ is the average of $X$.  We assume that the each Gaussian process $h_k(\tau, t)$ is independent and $h_k(\tau, t)$, $\psi_k$, $\tau_k$ and $b_{k,n}$ are independent. Then, SNR of the user $i$ is defined as
\begin{equation}
\operatorname{SNR}_i = \sqrt{\frac{\operatorname{Var}\{D_i\} }{\operatorname{Var}\{F_i\}  + \operatorname{Var}\{I_i\}  + \operatorname{Var}\{N_i\}}}.
\label{eq:SNR_def}
\end{equation}
In this section, we focus on the estimation of the lower bound of Eq. (\ref{eq:SNR_def}) under some assumptions.
It is known from \cite{pursley} and \cite{borth} that the variance of $N_i$ is
\begin{equation}
\operatorname{Var}\{N_i\} = \frac{1}{4}N_0T
\end{equation}
if $n(t)$ has a two-sided spectral density $\frac{1}{2}N_0$.\\
We make assumptions about the channel and Gaussian process $h_k(\tau,t)$ that
\begin{enumerate}
\item the Fourier transform of covariance function of $h_k(\tau, t)$ and its inverse Fourier transform exist. 
\item the channel is a wide-sense-stationary uncorrelated-scattering (WSSUS) channel \cite{bello}.
\item the channel is a frequency selective fading channel.
\item the Gaussian process $h_k(\tau,t)$ satisfies $h_k(\tau,t)=0$ when $\tau < 0$ \cite{variance} .
\item there is an integer $M_k$ that satisfies $h_k(\tau,t)=0$ when $\tau > M_kT$.
\end{enumerate}
In general, the faded signal is composed of the sum of the delayed signal which is affected by the Doppler shift and the delayed signals are attenuated as time passes. The models that the probability of time delay $\tau$ obeys an exponential distribution are often used \cite{peter}. The last assumption is equivalent to the one that the delayed signal gets to zero in finite-time. 
In WSSUS channels, the covariance function of $h_k(\tau, t)$ is expressed as \cite{borth}
\begin{equation}
\begin{split}
\Sigma_k(\tau_1, \tau_2; t_1, t_2) &= \operatorname{E}[h_k(\tau_1,t_1)\overline{h_k(\tau_2,t_2)}]\\
&=\rho_k(\tau_1, t_1-t_2)\delta(\tau_1 - \tau_2).
\end{split}
\end{equation}
Adding to this condition, in a selective fading channel, covariance function $\Sigma_k$ is \cite{borth}
\begin{equation}
\begin{split}
\Sigma_k(\tau_1, \tau_2; t_1, t_2) &= \rho_k(\tau_1, 0)\delta(\tau_1 - \tau_2)\\
&= g_k(\tau_1)\delta(\tau_1 - \tau_2).
\label{eq:covariance}
\end{split}
\end{equation}
In the above equation, we have defined $g_k(\tau) = \rho_k(\tau, 0)$. The covariance function $\Sigma_k$ is independent of $t_1$ and $t_2$. \\
We calculate the variance of $F_i$. With Eq. (\ref{eq:thm1}), $\operatorname{Var}\{F_i\}$ is
\begin{equation}
\begin{split}
\operatorname{Var}\{F_i\} &=\frac{P}{4}\operatorname{E}\{\operatorname{Re}[\tilde{F_i}^2]\} + \frac{P}{4}\operatorname{E}\{|\tilde{F_i}|^2\}.
\end{split}
\label{eq:var_f}
\end{equation}
Here, $\operatorname{E}\{\operatorname{Re}[\tilde{F_i}^2]\}$ and $\operatorname{E}\{|\tilde{F_i}|^2\}$ are expressed as
\begin{equation}
\begin{split}
\operatorname{E}\{\operatorname{Re}[\tilde{F_i}^2]\} =& \gamma_i^2 \cdot \operatorname{E}_{\mathbf{b}_i}\left\{ \operatorname{Re}\left[\int_0^T\int_0^T \int_{-\infty}^\infty \int_{-\infty}^\infty\tilde{\Sigma}_i(\tau_1,\tau_2;t_1,t_2)\right. \right. \\
&\left. \cdot \mu_{i}(\tau_1; t_1)\mu_{i}(\tau_2; t_2) d\tau_1 d\tau_2 dt_1 dt_2 \Bigg]\right\},\\
\operatorname{E}\{|\tilde{F_i}|^2\} =& \gamma_i^2 \cdot \operatorname{E}_{\mathbf{b}_i}\Bigg\{ \int_0^T\int_0^T \int_{-\infty}^\infty \int_{-\infty}^\infty\Sigma_i(\tau_1,\tau_2;t_1,t_2)\\
&\cdot \xi_i(\tau_1,\tau_2;t_1,t_2)d\tau_1 d\tau_2 dt_1 dt_2 \Bigg\},
\end{split}
\end{equation}
where 
\[\tilde{\Sigma}_i(\tau_1,\tau_2;t_1,t_2) = \operatorname{E}\left\{h_i(\tau_1,t_1)h_i(\tau_2,t_2)\right\}\]
and $\operatorname{E}_{\mathbf{b}_i}\{X\}$ is the average over all the bits of the user $i$. We write the variable over which we take average at the right bottom of $\operatorname{E}$. In \cite{ofdmandcdma}, it is shown that we can use
\begin{equation}
\operatorname{E}\{h_k(\tau_1,t_1)h_k(\tau_2,t_2)\} = 0.
\label{eq:rf}
\end{equation}
This result is obtained from the demodulation of RF signals.
From Eqs. (\ref{eq:covariance})-(\ref{eq:rf}), we have
\begin{equation}
\operatorname{Var}\{F_i\} = \frac{P}{4}\gamma_i^2  \cdot \operatorname{E}_{\mathbf{b}_i}\Bigg\{ \int_{-\infty}^\infty g_i(\tau) \int_0^T\int_0^T \xi_i(\tau, \tau ; t_1, t_2) dt_1dt_2d\tau \Bigg\}.
\label{eq:fad}
\end{equation}
The double integral term is written as
\begin{equation}
\begin{split}
\int_0^T\int_0^T \xi_i(\tau, \tau ; t_1, t_2)dt_1dt_2 &= \left(\int_0^T \mu_i(\tau; t_1) dt_1\right)\left(\int_0^T \overline{\mu_i(\tau; t_2)} dt_2\right)\\
&= \left|\int_0^T \mu_i(\tau; t) dt\right|^2 = \Gamma_i(\tau) \geq 0,
\end{split}
\label{eq:gamma_i}
\end{equation}
where $\Gamma_i(\tau)$ has been defined. Note that $\Gamma_i(\tau)$ is the squared absolute value of the correlation in an asynchronous CDMA system.
From the assumptions 4 and 5, we obtain 
\begin{equation}
\begin{split}
g_i(\tau) &= 0 \hspace{3mm}\mbox{for} \hspace{2mm}\tau < 0\\
g_i(\tau) &= 0 \hspace{3mm}\mbox{for} \hspace{2mm}\tau > M_iT.
\end{split}
\end{equation}
Note that $g_i(\tau)$ is non-negative since
\[g_i(\tau) = \rho_i(\tau,0) = \operatorname{E}\{h_i(\tau,t) \overline{h_i(\tau,t)}\}\geq0.\]
Further, we can assume that $g_i(\tau)$ has the upper bound $C_i$ in $[0,M_iT]$. This assumption is obtained from the assumption 1. We have no knowledge about the form of $g_i(\tau)$. For this reason, we evaluate the upper bound of $\operatorname{Var}\{F_i\}$ with the product of two terms, one is related to $g_i(\tau)$ and the other is related to the spreading sequences. From H\"older's inequality, we evaluate Eq. (\ref{eq:fad}) as
\begin{equation}
\begin{split}
\operatorname{Var}\{F_i\} =& \frac{P}{4}\gamma_i^2 \cdot \operatorname{E}_{\mathbf{b}_i}\left\{ \int_{-\infty}^\infty g_i(\tau) \Gamma_i(\tau) d\tau\right\} \\
\leq& \frac{P}{4}\gamma_i^2 \cdot \operatorname{E}_{\mathbf{b}_i}\left\{ \sup_{[0,M_iT]}\{|g_i(\tau)|\} \cdot \int_{0}^{M_iT}  |\Gamma_i(\tau)| d\tau \right\}\\
=&  \frac{P}{4}\gamma_i^2 C_i \cdot \operatorname{E}_{\mathbf{b}_i}\left\{ \int_{0}^{M_iT}  \Gamma_i(\tau) d\tau \right\}.
\end{split}
\label{eq:var_f}
\end{equation}
The equality is attained if $g_i(\tau)$ is the rectangular function. This is the worst case where $\operatorname{Var}\{F_i\}$ is maximized. We assume that the time delay $\tau$ satisfies $n_i T + l_i T_c \leq \tau < n_iT + (l_i + 1)T_c$, where $0 \leq n_i < M_i$ and $0 \leq l_i < N$ is an integer. Note that $n_i T + NT_c = (n_i + 1)T$. Since the correlation in an asynchronous CDMA system is the superposition of the correlations in a chip-synchronous CDMA system, the function $\Gamma_i(\tau)$ in Eq. (\ref{eq:gamma_i}) can be written as
 \begin{equation}
\begin{split}
\Gamma_i(\tau) &= \left|R_i\left(\tau, n_i, l_i\right) + \hat{R}_i\left(\tau, n_i, l_i\right) \right|^2,
\end{split}
\end{equation}
where
 \begin{equation}
 \begin{split}
 R_i\left(\tau, n, l \right) =& \left(\tau - nT - lT_c\right)\\
&\cdot \left\{b_{i,-n-1} \sum_{m=1}^{l}\overline{s_{i,m}}s_{i,N-l+m}+ b_{i,-n}\sum_{m=1}^{N-l} \overline{s_{i,l+m}}s_{i,m} \right\},\\
 \hat{R}_i\left(\tau, n, l\right) =& \left(nT + (l+1)T_c - \tau\right)\\
&\cdot \left\{b_{i,-n-1} \sum_{m=1}^{l+1}\overline{s_{i,m}}s_{i,N-l+m-1}+ b_{i,-n}\sum_{m=1}^{N-l-1} \overline{s_{i,l+m+1}}s_{i,m} \right\}.
 \end{split}
 \label{eq:cor}
 \end{equation}
 Note that $R_i\left(\tau, n, l \right)$ and $\hat{R}_i\left(\tau, n, l\right)$ are autocorrelation functions in a chip-synchronous CDMA model.
From Eq. (\ref{eq:cor}), it is sufficient to consider only two adjacent bits, $b_{i, -n_i-1}$ and $b_{i, -n_i}$. From the independence of each bit $b_{i,-n_i}$, Eq. (\ref{eq:var_f}) can be written as
 \begin{equation}
 \begin{split}
 \operatorname{Var}\{F_i\} &\leq \frac{P}{4}\gamma_i^2 C_i M_i \cdot \operatorname{E}_{\mathbf{b}_i}\left\{ \sum^{N-1}_{l_i=0}\int_{l_i T_c}^{(l_i + 1)T_c} \Gamma_i(\tau,0,l_i) d\tau\right\},
 \end{split}
 \label{eq:fad_cor}
 \end{equation}
 where
 \begin{equation*}
 \Gamma_i(\tau,n,l) = \left|R_i\left(\tau, n, l\right) + \hat{R}_i\left(\tau, n, l\right) \right|^2.
 \end{equation*}
Since $P\gamma_i^2 C_i M_i$ is a constant, it is sufficient to focus on the sum term in the right hand side of Eq. (\ref{eq:fad_cor}) to reduce the upper bound of $ \operatorname{Var}\{F_i\}$.

Similar to the fading term, we evaluate the interference noise term $I_i$. The variance of $I_i$ is
\begin{equation}
\operatorname{Var}\{I_i\} =\frac{P}{4}\sum^K_{\substack{k=1 \\ k \neq i}}\left[\gamma_k^2 \operatorname{Var}\{\tilde{I}_{i,k}\} + \operatorname{Var}\{\tilde{I}'_{i,k}\}\right],
\end{equation}
 where we have used Eqs. (\ref{eq:thm1}), (\ref{eq:rf}), and
 \begin{equation}
 \operatorname{E}_{\psi_k}\{\exp(2 j \psi_k)\} = 0.
 \end{equation}
Some components $\tilde{I}_{i,k}$ appeared in Eq. (\ref{eq:signals}) is the fading interference noise term and $\tilde{I}'_{i,k}$ is the term of the direct wave. With Eq. (\ref{eq:covariance}), variances of them are expressed as
\begin{equation}
\begin{split}
\operatorname{Var}\{\tilde{I}_{i,k}\}=&\operatorname{E}_{\mathbf{b}_k,\tau_k}\left\{\int_{-\infty}^{\infty} \int_0^T \int_0^T  g_k(\tau) \right. \\
 \cdot & \xi_{i,k,k}(\tau+\tau_k, \tau+\tau_k; t_1,t_2)dt_1dt_2 d\tau \biggr\},\\
\operatorname{Var}\{\tilde{I}'_{i,k}\}=& \operatorname{E}_{\mathbf{b}_k,\tau_k}\left\{\int_0^T \int_0^T  \xi_{i,k,k}(\tau_k, \tau_k; t_1,t_2) dt_1dt_2\right\}.
\end{split}
\end{equation}
We assume that $l_k T_c \leq \tau_k < (l_k + 1)T_c$, where $0 \leq l_k < N$ is an integer. The above double integral term is written as
\begin{equation}
\begin{split}
&\int_0^T \int_0^T  \xi_{i,k,k}(\tau_k, \tau_k; t_1,t_2) dt_1dt_2\\
=& \left|R_{i,k}\left(\tau_k ,0, l_k\right) + \hat{R}_{i,k}\left(\tau_k, 0, l_k\right) \right|^2,
\end{split}
\end{equation}
where
 \begin{equation}
 \begin{split}
 R_{i,k}\left(\tau, n, l \right) =& \left(\tau - nT - lT_c\right)\\
\cdot& \left\{b_{k,-n-1} \sum_{m=1}^{l}\overline{s_{i,m}}s_{k,N-l+m}+ b_{k,-n}\sum_{m=1}^{N-l} \overline{s_{i,l+m}}s_{k,m} \right\},\\
 \hat{R}_{i,k}\left(\tau, n, l\right) =& \left(nT + (l+1)T_c - \tau\right)\\
\cdot& \left\{b_{k,-n-1} \sum_{m=1}^{l+1}\overline{s_{i,m}}s_{i,N-l+m-1}+ b_{k,-n}\sum_{m=1}^{N-l-1} \overline{s_{i,l+m+1}}s_{k,m} \right\}.
 \end{split}
 \label{eq:cor2}
 \end{equation}
 Note that $R_{i,k}\left(\tau, n, l \right)$ and $\hat{R}_{i,k}\left(\tau, n, l\right)$ are crosscorrelation functions in a chip-synchronous CDMA model.
 We define
  \begin{equation}
 \Gamma_{i,k}(\tau,n,l) = \left|R_{i,k}\left(\tau ,n, l\right) + \hat{R}_{i,k}\left(\tau, n, l\right) \right|^2,
   \end{equation}
so that $\operatorname{Var}\{\tilde{I}'_{i,k}\}$ is concisely written as 
\begin{equation}
\begin{split}
\operatorname{Var}\{\tilde{I}'_{i,k}\}&= \frac{1}{T} \cdot \operatorname{E}_{\mathbf{b}_k}\left\{ \sum_{l_k=0}^{N-1}\int_{l_kT_c}^{(l_k+1)T_c}  \Gamma_{i,k}(\tau_k,0,l_k) d\tau_k \right\}.
\end{split}
\label{eq:tilde_I}
\end{equation}
 We consider the variance of $\tilde{I}_{i,k}$. Similar to the fading signal term, we assume that $n'_k T + l'_k T_c \leq \tau < n'_kT + (l'_k + 1)T_c$, where $0 \leq l'_k < N$ is an integer and $n'_k \geq 0$ is an integer. When we take the average over the bits, since each bit $b_{k,-n}$ is independent, it is sufficient to consider only the two adjacent bits $b_{k,-n'_k}$ and $b_{k,-n'_k-1}$. From these properties, we obtain
\begin{equation}
 \begin{split}
  &\frac{1}{T}\cdot\operatorname{E}_{\mathbf{b}_k}\left\{\int^T_0\int_0^T \int_0^T \xi_{i,k,k}(\tau+\tau_k, \tau+\tau_k; t_1,t_2)dt_1dt_2 d\tau_k \right\}\\
 =&\frac{1}{T} \cdot\operatorname{E}_{\mathbf{b}_k}\left\{\sum_{l_k=0}^{N-1}\int_{l_kT_c}^{(l_k+1)T_c} \Gamma_{i,k}\left(\tau_k, n'_k, l_k\right) d\tau_k \right\}\\
 =&\operatorname{Var}\{\tilde{I}'_{i,k}\}.
   \end{split}
\end{equation}
In the above equations, we have set $n'_k = 0$ to obtain the last equality. Then, we can express $\operatorname{Var}\{\tilde{I}_{i,k}\}$ as the product of $\operatorname{Var}\{\tilde{I}'_{i,k}\}$ and the integral covariance term.
With the above results, we have
\begin{equation}
\operatorname{Var}\{\tilde{I}_{i,k}\} = \frac{1}{T}L_k \cdot\operatorname{E}_{\mathbf{b}_k}\left\{\sum_{l_k=0}^{N-1}\int_{l_kT_c}^{(l_k+1)T_c}\Gamma_{i,k} \left(\tau_k,0, l_k\right) d\tau_k \right\},
\end{equation}
where
\begin{equation}
L_k = \int_{-\infty}^{\infty}g_k(\tau)d \tau = \int_{0}^{M_kT}g_k(\tau)d \tau.
\end{equation}
In the worst case for $\operatorname{Var}\{F_i\}$, where $g_i(\tau)$ is the rectangular function, $L_k$ is
\begin{equation}
L_k = M_kC_kT.
\end{equation}
From these calculations, the variance of $I_i$ is
\begin{equation}
\begin{split}
\operatorname{Var}\{I_{i}\} =& \frac{P}{4T}\sum^K_{\substack{k=1 \\ k \neq i}} (1 + \gamma^2_kL_k) \cdot\operatorname{E}_{\mathbf{b}_k} \left\{ \sum_{l_k=0}^{N-1} \int_{l_kT_c}^{(l_k+1)T_c} \Gamma_{i,k}\left(\tau_k, 0, l_k\right)  d\tau_k \right\}.
\end{split}
\label{eq:inter_cor}
\end{equation}
To increase the lower bound of SNR, it is necessary and sufficiently to reduce the sum and integral term since $(1 + \gamma^2_kL_k)$ is constant.

\section{New Expression of SNR formula}
First, we consider the interference noise term. In \cite{basis}, it is shown that the crosscorrelation of chip-synchronous CDMA system can be written in the quadratic form. With this expression, we can rewrite $\Gamma_{i,k}(\tau_k,0,l_k)$ in Eq. (\ref{eq:inter_cor}) as
 \begin{equation}
  \begin{split}
   &\Gamma_{i,k}(\tau_k,0,l_k) \\
=& \left| \left(\tau_k - l_kT_c\right) \mathbf{s}^*_i B^{(l_k)}_{b_{k,-1},b_{k,0}} \mathbf{s}_k \right. \left.+  \left((l_k+1)T_c - \tau_k\right)\mathbf{s}^*_i B^{(l_k+1)}_{b_{k,-1},b_{k,0}} \mathbf{s}_k \right|^2\\
 =& \left(\tau_k - l_kT_c\right)^2\left|\mathbf{s}^*_i B^{(l_k)}_{b_{k,-1},b_{k,0}} \mathbf{s}_k\right|^2 + \left((l_k+1)T_c - \tau_k\right)^2\left|\mathbf{s}^*_i B^{(l_k+1)}_{b_{k,-1},b_{k,0}} \mathbf{s}_k \right|^2\\
 +&  2\left(\tau_k - l_kT_c\right) \left((l_k+1)T_c - \tau_k\right)\operatorname{Re}\left[\left(\mathbf{s}^*_i B^{(l_k)}_{b_{k,-1},b_{k,0}} \mathbf{s}_k\right) \overline{\left(\mathbf{s}^*_i B^{(l_k+1)}_{b_{k,-1},b_{k,0}} \mathbf{s}_k\right)} \right],
   \end{split}
 \end{equation}
where $\mathbf{z}^*$ is a complex conjugate transpose of $\mathbf{z}$,
\begin{equation}
\mathbf{s}_k = (s_{k,1}, s_{k,2}, \ldots, s_{k,N})^\mathrm{T}
\end{equation}
and
\begin{equation}
B^{(l)}_{b_{k,-1},b_{k,0}} = \left( \begin{array}{c c}
O & b_{k,-1}E_{l} \\
b_{k,0}E_{N-l} & O
\end{array}
\right).
\end{equation}
In the above equations, $\mathbf{s}^\mathrm{T}$ is the transpose of $\mathbf{s}$ and $E_l$ is the identity matrix of size $l$. It can be shown that $\mathbf{s}_k$ is expressed as \cite{basis}
\begin{equation}
\mathbf{s}_k = \frac{1}{\sqrt{N}}\sum_{m=1}^N \alpha^{(k)}_m\mathbf{w}_m(0)=\frac{1}{\sqrt{N}}\sum_{m=1}^N \beta^{(k)}_m\mathbf{w}_m\left(\frac{1}{2N}\right),
\end{equation}
where $\mathbf{w}_m(\eta)$ is the basis vector whose $n$-th component is expressed as
\[\left(\mathbf{w}_m(\eta)\right)_n = \exp\left(2\pi j (n-1)\left(\frac{m}{N} + \eta\right)\right).\]
When we calculate the integral of $\Gamma_{i,k}(\tau_k,0,l_k)$, we have 
\begin{equation}
\begin{split}
&\int_{l_kT_c}^{(l_k+1)T_c} \Gamma_{i,k}(\tau_k,0,l_k) d\tau_k\\
 =& \frac{1}{3}T_c^3\left|\mathbf{s}^*_i B^{(l_k)}_{b_{k,-1},b_{k,0}} \mathbf{s}_k\right|^2 + \frac{1}{3}T_c^3\left|\mathbf{s}^*_i B^{(l_k+1)}_{b_{k,-1},b_{k,0}} \mathbf{s}_k \right|^2\\
 +& \frac{1}{3}T_c^3 \operatorname{Re}\left[\left(\mathbf{s}^*_i B^{(l_k)}_{b_{k,-1},b_{k,0}} \mathbf{s}_k\right) \overline{\left(\mathbf{s}^*_i B^{(l_k+1)}_{b_{k,-1},b_{k,0}} \mathbf{s}_k\right)} \right].
\end{split}
\label{eq:inter_term}
\end{equation}
When we take the average of Eq. (\ref{eq:inter_term}) over the bits $b_{k,-1}$ and $b_{k,0}$, the averaged quantity is
\begin{equation}
\begin{split}
\operatorname{E}_{\mathbf{b}_k}\left\{\left|\mathbf{s}^*_i B^{(l_k)}_{b_{k,-1},b_{k,0}} \mathbf{s}_k\right|^2\right\} =& \frac{1}{2}\left\{ \left| \sum_{m=1}^{N}\lambda_m^{(l_k)} \overline{\alpha^{(i)}_m}\alpha^{(k)}_m\right|^2 + \left| \sum_{m=1}^{N}\hat{\lambda}_m^{(l_k)} \overline{\beta^{(i)}_m}\beta^{(k)}_m\right|^2\right\},
\end{split}
\end{equation}
and
\begin{equation}
\begin{split}
&\operatorname{E}_{\mathbf{b}_k}\left\{\operatorname{Re}\left[\left(\mathbf{s}^*_i B^{(l_k)}_{b_{k,-1},b_{k,0}} \mathbf{s}_k\right) \overline{\left(\mathbf{s}^*_i B^{(l_k+1)}_{b_{k,-1},b_{k,0}} \mathbf{s}_k\right)} \right]\right\}\\
=&\frac{1}{2} \operatorname{Re}\left[ \left(\sum_{m=1}^{N}\lambda_m^{(l_k)} \overline{\alpha^{(i)}_m}\alpha^{(k)}_m\right) \overline{\left(\sum_{m'=1}^{N}\lambda_{m'}^{(l_k+1)} \overline{\alpha^{(i)}_{m'}}\alpha^{(k)}_{m'}\right)}\right] \\
+& \frac{1}{2} \operatorname{Re}\left[ \left(\sum_{m=1}^{N}\hat{\lambda}_m^{(l_k)} \overline{\beta^{(i)}_m}\beta^{(k)}_m\right) \overline{\left(\sum_{m'=1}^{N}\hat{\lambda}_{m'}^{(l_k+1)} \overline{\beta^{(i)}_{m'}}\beta^{(k)}_{m'}\right)}\right],
\end{split}
\end{equation}
where $\lambda_m^{(l)} =  \exp\left(-2 \pi j l\frac{m}{N}\right)$ and $\hat{\lambda}_m^{(l)} =  \exp\left(-2 \pi j l\left(\frac{m}{N} + \frac{1}{2N}\right)\right)$.
When we take the sum over $l_k$, we rewrite Eq. (\ref{eq:inter_cor}) as
\begin{equation}
\operatorname{Var}\{I_{i,k}\} = \frac{PT^2}{12N^2}\sum_{\substack{k=1 \\ k \neq i}}^K (1 + \gamma_k^2 L_k)\sum_{m=1}^N S^{i,k}_m,
\end{equation}
where
\begin{equation}
\begin{split}
S^{i,k}_m &= \left|\alpha_m^{(i)}\right|^2\left|\alpha_m^{(k)}\right|^2\left(1 + \frac{1}{2}\cos\left(2 \pi \frac{m}{N}\right)\right)\\
&+ \left|\beta_m^{(i)}\right|^2\left|\beta_m^{(k)}\right|^2\left(1 + \frac{1}{2}\cos\left(2 \pi \left(\frac{m}{N} + \frac{1}{2N}\right)\right)\right).
\end{split}
\end{equation}
When we replace $k$ with $i$, we obtain the expression of $\Gamma_i(\tau,0,l_i)$. Then, Eq. (\ref{eq:fad_cor}) is rewritten as
\begin{equation}
\operatorname{Var}\{F_i\} \leq \frac{PT^3}{12N^2}\gamma_i^2 C_i M_i \sum_{m=1}^N S_m^{i,i}.
\end{equation}
From the above expressions, we arrive at the formula for the lower bound of SNR
\begin{equation}
\operatorname{SNR}_i \geq \left\{\frac{1}{6N^2}\sum_{k=1}^K Z_{i,k} \sum_{m=1}^N S_m^{i,k} + \frac{N_0}{2PT}\right\}^{-1/2},
\label{eq:SNR}
\end{equation}
where
\[ Z_{i,k} = \left\{ \begin{array}{c c}
\displaystyle \gamma_i^2 C_i M_iT & i=k\\
\displaystyle 1 + \gamma_k^2 L_k & i \neq k
\end{array}
\right. .\]

\section{Optimization Problem for SNR}
Our goal is to derive necessary conditions for the optimal spreading sequence which maximizes SNR for the case where the $g_i(\tau)$ is the worst. Here, the condition that $Z_{i,k}$ is fixed is equivalent to that $\gamma_k$, $M_k$, $C_k$, $N$, $K$ and $T$ are fixed. We treat $Z_{i,k}$ as the weights among all the users. We ignore the Gaussian noise term since it has no relation to spreading sequences. To maximize the lower bound of SNR of the user $i$, we should minimize the first term of the denominator of Eq. (\ref{eq:SNR}). We consider the optimization problem $(\tilde{P})$
\begin{equation}
\begin{split}
(\tilde{P}) & \hspace{3mm} \min \hspace{2mm}\sum_{k=1}^K Z_{i,k} \sum_{m=1}^N S_m^{i,k}\\
\mbox{subject to}  \hspace{3mm} & {\boldsymbol \alpha^{(k)}} = \Phi {\boldsymbol \beta^{(k)}} \hspace{3mm}(k=1,2,\ldots,K),\\
& {\boldsymbol \beta^{(k)}} = \hat{\Phi} {\boldsymbol \alpha^{(k)}} \hspace{3mm}(k=1,2,\ldots,K),\\
&\left\|\boldsymbol \alpha^{(k)}\right\|^2 = \left\|\boldsymbol \beta^{(k)}\right\|^2 = N \hspace{3mm}(k=1,2,\ldots,K),
\end{split}
\end{equation}
where $\|\mathbf{x}\|$ is the Euclidian norm of the vector $\mathbf{x}$,
\begin{equation}
 {\boldsymbol \alpha^{(k)}} = \left( \begin{array}{c}
 \alpha_1^{(k)}\\
 \alpha_2^{(k)}\\
 \vdots\\
 \alpha_N^{(k)}
 \end{array} \right),  {\boldsymbol \beta^{(k)}} = \left( \begin{array}{c}
 \beta_1^{(k)}\\
 \beta_2^{(k)}\\
 \vdots\\
 \beta_N^{(k)}
 \end{array} \right),
  \end{equation}
  $\Phi$ and $\hat{\Phi}$ are the unitary matrices whose $(m,n)$-th components are
  \begin{equation}
  \begin{split}
   \Phi_{m,n}=&\frac{1}{N}\cdot\frac{2}{1-\exp(2 \pi j (\frac{n-m}{N} + \frac{1}{2N}))},\\
 \hat{\Phi}_{m,n}  =&\frac{1}{N}\cdot\frac{2}{1-\exp(2 \pi j (\frac{n-m}{N} - \frac{1}{2N}))}.
  \end{split}
  \end{equation}
Note that the objective function is a convex function and this problem is non-linear convex problem if we did not take into account the norm constraints. The first two constraint terms have been discussed in \cite{basis}. The last constraint term is a signal power constraint.\\
We should take into account all the users for designing spreading sequences. It might be appropriate that we maximize the sum of $\operatorname{SNR}_i$. However, it is difficult to obtain the global solution since the objective function is not convex and the derivative is complicated. Therefore, we consider the problem $(\tilde{P}')$ which consists of the sum of the denominator of $\operatorname{SNR}_i$
\begin{equation}
\begin{split}
(\tilde{P}')& \hspace{3mm} \min \hspace{2mm}\sum_{i=1}^K\sum_{k=1}^K Z_{i,k} \sum_{m=1}^N S_m^{i,k}\\
\mbox{subject to}  \hspace{3mm} & {\boldsymbol \alpha^{(k)}} = \Phi {\boldsymbol \beta^{(k)}} \hspace{3mm}(k=1,2,\ldots,K),\\
& {\boldsymbol \beta^{(k)}} = \hat{\Phi} {\boldsymbol \alpha^{(k)}} \hspace{3mm}(k=1,2,\ldots,K),\\
&\left\|\boldsymbol \alpha^{(k)}\right\|^2 = \left\|\boldsymbol \beta^{(k)}\right\|^2 = N \hspace{3mm}(k=1,2,\ldots,K).
\end{split}
\end{equation}
The objective function of the problem $(\tilde{P}')$ is convex. This problem is the natural extension of the problem $(\tilde{P})$. There is the relation between the sum of the denominator of $\operatorname{SNR}_i$ and the sum of $\operatorname{SNR}_i$ as
\begin{equation}
\begin{split}
\frac{1}{K}\sum_{i=1}^K \operatorname{SNR}_i  & \geq \left[\frac{1}{\sum_{i=1}^K \{\operatorname{Denom}(\operatorname{SNR}_i )\}^2}\right]^{-1/2},
\end{split}
\end{equation}
where $\operatorname{Denom}(\operatorname{SNR}_i )$ is the denominator of $\operatorname{SNR}_i$. In the problem $(\tilde{P}')$, we evaluate the lower bound of the sum of $\operatorname{SNR}_i$.\\
The variables in the problem $(\tilde{P}')$ is complex numbers. We rewrite the problem $(\tilde{P}')$ to the real number optimization problem. It is shown \cite{cmgc} the method of transforming a complex-number vector to a real-number vector and a complex-number unitary matrix to a real-number orthogonal matrix. With this result, we consider the problem $(P)$
\begin{equation}
\begin{split}
(P)& \hspace{3mm} \min \hspace{2mm}\sum_{i=1}^K\sum_{k=1}^K Z_{i,k} \sum_{m=1}^N \hat{S}_m^{i,k}\\
\mbox{subject to}  \hspace{3mm} & {\boldsymbol \alpha'^{(k)}} = \Phi' {\boldsymbol \beta'^{(k)}} \hspace{3mm}(k=1,2,\ldots,K),\\
& {\boldsymbol \beta'^{(k)}} = \hat{\Phi}' {\boldsymbol \alpha'^{(k)}} \hspace{3mm}(k=1,2,\ldots,K),\\
&\left\|\boldsymbol \alpha'^{(k)}\right\|^2 = \left\|\boldsymbol \beta'^{(k)}\right\|^2 = N \hspace{3mm}(k=1,2,\ldots,K),
\end{split}
\end{equation}
where
\begin{equation}
\begin{split}
\Phi' &= \left( \begin{array}{c c}
\operatorname{Re}[\Phi] & -\operatorname{Im}[\Phi]\\
\operatorname{Im}[\Phi] & \operatorname{Re}[\Phi]
\end{array} \right), \hat{\Phi}' = \left( \begin{array}{c c}
\operatorname{Re}[\hat{\Phi}] & -\operatorname{Im}[\hat{\Phi}]\\
\operatorname{Im}[\hat{\Phi}] & \operatorname{Re}[\hat{\Phi}]
\end{array} \right), \\
{\boldsymbol \alpha}'^{(k)} &= 
\left( \begin{array}{c}
{\boldsymbol \alpha}^{(k)} _{1}\\
{\boldsymbol \alpha}^{(k)} _{2}
\end{array} \right)
=\left( \begin{array}{c}
\operatorname{Re}[{\boldsymbol \alpha}^{(k)}]\\
\operatorname{Im}[{\boldsymbol \alpha}^{(k)}]
\end{array} \right), {\boldsymbol \beta}'^{(k)}  = 
\left( \begin{array}{c}
{\boldsymbol \beta}^{(k)} _{1}\\
{\boldsymbol \beta}^{(k)} _{2}
\end{array} \right) = 
\left( \begin{array}{c}
\operatorname{Re}[{\boldsymbol \beta}^{(k)}]\\
\operatorname{Im}[{\boldsymbol \beta}^{(k)}]
\end{array} \right)
\end{split}
\end{equation}
and
\begin{equation}
\begin{split}
\hat{S}_m^{i,k}&= \left(\left(\alpha^{(i)}_{1,m}\right)^2 +\left(\alpha^{(i)}_{2,m}\right)^2  \right) \left(\left(\alpha^{(k)}_{1,m}\right)^2 +\left(\alpha^{(k)}_{2,m}\right)^2  \right) \left(1 + \frac{1}{2}\cos\left(2\pi\frac{m}{N}\right)\right)\\
&+\left(\left(\beta^{(i)}_{1,m}\right)^2 +\left(\beta^{(i)}_{2,m}\right)^2  \right)\left(\left(\beta^{(k)}_{1,m}\right)^2 +\left(\beta^{(k)}_{2,m}\right)^2  \right)\\
&\cdot \left(1 + \frac{1}{2}\cos\left(2\pi\left(\frac{m}{N} + \frac{1}{2N}\right)\right)\right).
 \end{split}
\end{equation}
$\alpha^{(k)}_{1,m}$, $\alpha^{(k)}_{2,m}$, $\beta^{(k)}_{1,m}$ and $\beta^{(k)}_{2,m}$ are the $m$-th elements of ${\boldsymbol \alpha}^{(k)}_{1}$, ${\boldsymbol \alpha}^{(k)}_{2}$, ${\boldsymbol \beta}^{(k)}_{1}$ and ${\boldsymbol \beta}^{(k)}_{2}$.
We can reduce the two linear constraints to the one constraint, and the two norm constraints to the one norm constraint since $\hat{\Phi}'$ is an orthogonal matrix. From these reductions, we obtain the problem $(P')$
\begin{equation}
\begin{split}
(P') &\hspace{3mm} \min \hspace{2mm}\sum_{i=1}^K\sum_{k=1}^K Z_{i,k} \sum_{m=1}^N \hat{S}_m^{i,k}\\
\mbox{subject to}  \hspace{3mm} & {\boldsymbol \beta'^{(k)}} = \hat{\Phi}' {\boldsymbol \alpha'^{(k)}} \hspace{3mm}(k=1,2,\ldots,K),\\
&\left\|\boldsymbol \alpha'^{(k)}\right\|^2 = N \hspace{3mm}(k=1,2,\ldots,K),
\end{split}
\end{equation}
We collect the variables of the problem $(P')$ into $\mathbf{x}$. Here, $\mathbf{x}$ is expressed as
\begin{equation}
\mathbf{x} = \left( \begin{array}{c} 
\mathbf{x}_1\\
\mathbf{x}_2\\
\vdots\\
\mathbf{x}_K
\end{array} \right), \mathbf{x}_k = \left( \begin{array}{c}
\boldsymbol\alpha'^{(k)}\\
\boldsymbol\beta'^{(k)}
\end{array} \right).
\end{equation}
The dimension of $\mathbf{x}$ is $4NK$. The problem $(P')$ is a non-linear programming. There are many numerically methods for solving non-linear programmings.

\section{KKT Conditions for Optimized Spreading Sequences}
We show the necessary conditions for the global solution of problem $(P')$. To this end, we focus on the KKT conditions, where such conditions are the necessary conditions for the global solutions \cite{KKT}. \\
To differentiate the objective function and the constraint functions of the problem $(P')$, we express them as
\begin{equation}
\begin{split}
f(\mathbf{x}) &= \sum_{i=1}^K\sum_{k=1}^K Z_{i,k} \sum_{m=1}^N \hat{S}_m^{i,k},\\
c^{(k)}(\mathbf{x}) &=  {\boldsymbol \beta'^{(k)}} - \hat{\Phi}' {\boldsymbol \alpha'^{(k)}}\hspace{3mm} (k=1,2,\ldots,K),\\
d^{(k)}(\mathbf{x}) &= \left\|\boldsymbol \alpha'^{(k)}\right\|^2 - N\hspace{3mm} (k=1,2,\ldots,K).
\end{split}
\end{equation}
The constraint function $c^{(k)}(\mathbf{x})$ is a vector valued function. We divide the condition $c^{(k)}(\mathbf{x})$ into $2N$ conditions
\begin{equation}
\begin{split}
c^{(k)}_{1,m}(\mathbf{x}) &= \beta^{(k)}_{1,m} - \frac{1}{N}\left\{\sum_{n=1}^N \alpha^{(k)}_{1,n} - \sum_{n=1}^N \alpha^{(k)}_{2,n}\hat{\phi}_{m,n}\right\}, \\
c^{(k)}_{2,m}(\mathbf{x}) &= \beta^{(k)}_{2,m} - \frac{1}{N} \left\{ \sum_{n=1}^N \alpha^{(k)}_{1,n}\hat{\phi}_{m,n}  + \sum_{n=1}^N \alpha^{(k)}_{2,n} \right\},
\end{split}
\end{equation}
for $m=1,2,\dots,N$. In the above equations, $\hat{\phi}_{m,n}$ is 
\begin{equation}
\hat{\phi}_{m,n} = \frac{\sin\left(2 \pi \left(\frac{n-m}{N} - \frac{1}{2N}\right)\right)}{1- \cos\left(2 \pi \left(\frac{n-m}{N} - \frac{1}{2N}\right)\right)}
\end{equation}
and we have used
\begin{equation}
\begin{split}
\operatorname{Re}[\hat{\Phi}_{m,n}] &=\frac{1}{N},\\
\operatorname{Im}[\hat{\Phi}_{m,n}] &= \frac{1}{N} \cdot \frac{\sin\left(2 \pi \left(\frac{n-m}{N} - \frac{1}{2N}\right)\right)}{1- \cos\left(2 \pi \left(\frac{n-m}{N} - \frac{1}{2N}\right)\right)}.\\
\end{split}
\end{equation}
We focus on the user $p$. For the user $p$, $c^{(k)}_{1,m}(\mathbf{x})$, $c^{(k)}_{2,m}(\mathbf{x})$ and $d^{(k)}(\mathbf{x})$ have no relation when $k \neq p$. It is sufficient to consider only $c^{(p)}_{1,m}(\mathbf{x})$, $c^{(p)}_{2,m}(\mathbf{x})$ and $d^{(p)}(\mathbf{x})$ as constraint functions. \\
We define $\lambda^{(k)}_{1,m}$, $\lambda^{(k)}_{2,m}$ and $\mu^{(k)}$ as the Lagrange multipliers for $c^{(k)}_{1,m}(\mathbf{x})$, $c^{(k)}_{2,m}(\mathbf{x})$ and $d^{(k)}(\mathbf{x})$. These variables satisfy
\begin{equation}
\begin{split}
&\nabla f(\mathbf{\tilde{x}}) + \sum_{k=1}^K \sum_{m=1}^N \left\{ \lambda^{(k)}_{1,m} \nabla c^{(k)}_{1,m}(\mathbf{\tilde{x}}) +  \lambda^{(k)}_{2,m}\nabla c^{(k)}_{2,m}(\mathbf{\tilde{x}})\right\}\\
&+   \sum_{k=1}^K \mu^{(k)} \nabla d^{(k)}(\mathbf{\tilde{x}}) = \mathbf{0},
\label{eq:KKT}
\end{split}
\end{equation}
where $\mathbf{\tilde{x}}$ is the global solution of the problem $(P')$. Equation (\ref{eq:KKT}) is the KKT conditions for the problem $(P')$ and is the necessary condition that $\mathbf{\tilde{x}}$ is the global solution. If it is the case, $\lambda^{(k)}_{1,m}$, $\lambda^{(k)}_{2,m}$ and $\mu^{(k)}$ are real numbers.\\
From Eq. (\ref{eq:KKT}), $\beta^{(p)}_{1,q}$ and $\beta^{(p)}_{2,q}$ must satisfy
\begin{equation}
\begin{split}
\lambda^{(p)}_{1,q} =& -2\beta^{(p)}_{1,q}\sum_{k=1}^K (Z_{k,p} + Z_{p,k}) B^{(k)}_q  ,\\
\lambda^{(p)}_{2,q} =& -2\beta^{(p)}_{2,q}\sum_{k=1}^K (Z_{k,p} + Z_{p,k}) B^{(k)}_q,
\end{split}
\end{equation}
where
\begin{equation}
\begin{split}
A^{(k)}_q &= \left\{ \left(\alpha^{(k)}_{1,q}\right)^2 + \left(\alpha^{(k)}_{2,q}\right)^2 \right\}\left(1 + \frac{1}{2}\cos\left(2 \pi \frac{q}{N}\right)\right),\\
B^{(k)}_q &= \left\{ \left(\beta^{(k)}_{1,q}\right)^2 + \left(\beta^{(k)}_{2,q}\right)^2 \right\}\left(1 + \frac{1}{2}\cos\left(2 \pi \left(\frac{q}{N} + \frac{1}{2N}\right)\right)\right).
\end{split}
\end{equation}
From the above equations, $\lambda^{(p)}_{1,q}$ and $\lambda^{(p)}_{2,q}$ are determined. Similar to $\beta^{(p)}_{1,q}$ and $\beta^{(p)}_{2,q}$, $\alpha^{(p)}_{1,q}$ and $\alpha^{(p)}_{2,q}$ must satisfy 
\begin{equation}
\begin{split}
&2N\alpha^{(p)}_{1,q}\left[ \mu^{(p)} + \sum_{k=1}^K (Z_{k,p} + Z_{p,k})A_q^{(k)} \right]\\
 =& \sum_{m=1}^N \lambda_{1,m}^{(p)} + \sum_{m=1}^N \lambda_{2,m}^{(p)} \hat{\phi}_{m,q},\\
&2N\alpha^{(p)}_{2,q}\left[ \mu^{(p)} +  \sum_{k=1}^K (Z_{k,p} + Z_{p,k})A_q^{(k)} \right]\\
 =& -\sum_{m=1}^N \lambda_{1,m}^{(p)} \hat{\phi}_{m,q} + \sum_{m=1}^N \lambda_{2,m}^{(p)}
\end{split}
\label{eq:necessary}
\end{equation}
for all $q$. The Lagrange multiplier $\mu^{(p)}$ is the common variables in the above $2N$ equations. Since  $\lambda^{(k)}_{1,m}$, $\lambda^{(k)}_{2,m}$ are given, it is necessary for the global solution $\mathbf{\tilde{x}}$ that $\mu^{(p)}$ exists which satisfies Eq. (\ref{eq:necessary}).

\section{Conclusion}
We show the new expression of SNR formula. With this expression, we evaluate the lower bound of SNR. For a frequency-selective WSSUS Rician fading channel, we obtain the optimization problem: maximize that the lower bound of SNR and its necessary conditions. This optimization problem gets to the convex programming when we did not take into account the norm constraints. Therefore, the global solution can be numerically obtained if we deal with the norm constraints.\\
The remaining issue is to obtain a clear problem which maximizes all the SNR. The capacity limit of CDMA is expected to be clear if this issue is solved.




%

\end{document}